\documentclass[aps,prx,reprint,superscriptaddress,longbibliography,nofootinbib]{revtex4-2}

\usepackage[T1]{fontenc}
\usepackage[utf8]{inputenc}
\usepackage{amsmath,amssymb,amsfonts,bm}
\usepackage{graphicx}
\usepackage{booktabs}
\usepackage{array}
\usepackage{microtype}
\usepackage[table]{xcolor}
\usepackage{url}
\usepackage[hidelinks]{hyperref}

\graphicspath{{}{Figures/Figures_NC/}{Figures/}}

\newcommand{\RC}{\mathrm{RC}}

\newcommand{\ic}[1]{#1}
\newcommand{\figbox}[2][]{%
  \IfFileExists{#2}{\includegraphics[#1]{#2}}{%
    \fbox{\begin{minipage}[c][0.25\textheight][c]{0.92\linewidth}
    \centering
    Figure file not found: \texttt{\detokenize{#2}}.\\
    Replace this box with the final production figure.
    \end{minipage}}}}

\begin{document}

% PRX Popular Summary, to be submitted as a separate nontechnical item.
% Mutations are usually described as many separate rates, one for each local DNA context. Here we ask whether those rates are organized by a simpler physical structure. By comparing every mutation with its reverse, human mutation probabilities define a directed field on DNA sequence space. We show that most of this field is described by an effective landscape, much like an energy landscape in equilibrium physics, while the remaining part forms irreversible circulation around closed mutation cycles. Remarkably, the landscape is learned only from mutation probabilities but recovers major features of genome composition and reverse-complement symmetry. The irreversible component is smaller, reproducible across populations, and enriched in CpG-associated transitions, a known methylation-related mutational process. The work provides a general way to turn mutation tables into interpretable maps of the equilibrium-like and non-equilibrium forces shaping sequence evolution.

\title{Human mutation field reveals an equilibrium-like structure with irreversible circulation}
%Landscape--curl decomposition of human mutation dynamics on sequence space}

\author{Isabella Caranzano}
\affiliation{AI and Computational Biomedicine Unit, Department of Medical Sciences, University of Torino, Torino, Italy}

\author{Daniel Maria Busiello}
\email{danielmaria.busiello@unipd.it}
\affiliation{Department of Physics, University of Padova, Padova, Italy}
\affiliation{Max Planck Institute for the Physics of Complex Systems, Dresden, Germany}

\author{Stefano Priorelli}
\affiliation{AI and Computational Biomedicine Unit, Department of Medical Sciences, University of Torino, Torino, Italy}

\author{Amos Maritan}
\affiliation{Department of Physics, University of Padova, Padova, Italy}

\author{Piero Fariselli}
\email{piero.fariselli@unito.it}
\affiliation{AI and Computational Biomedicine Unit, Department of Medical Sciences, University of Torino, Torino, Italy}

\date{\today}

\begin{abstract}
The evolution of DNA sequences can be viewed as stochastic dynamics on a high-dimensional discrete space, but it is unclear when empirical transition biases can be reduced to an effective energy landscape and when irreducible non-equilibrium circulation remains. Human context-dependent mutation probabilities provide a direct test: every single-nucleotide substitution in a local sequence context has a reverse substitution. Therefore, the logarithm of the forward-to-reverse probability ratio defines an antisymmetric field that captures human mutation biases, here termed the human mutation field. Here we show that this field admits a dominant gradient component and a smaller but reproducible curl component. Using seven-base human germline substitution probabilities, we infer an effective mutational landscape with a Siamese neural architecture constrained to predict only energy differences. This model predicts forward-to-reverse log-ratios for mutations excluded from training with a correlation of about 0.93, close to both an unconstrained predictive reference (0.948) and the empirical reversible ceiling obtained by Hodge projection (about 0.96). Although trained only on mutation probabilities, the inferred landscape substantially recovers short-word genomic composition and Chargaff reverse-complement symmetry for sequences up to length four. Deviations from an equilibrium structure constitute the signature of a small, but detectable, nonequilibrium component. This is associated with a residual irreversible circulation that violates the Kolmogorov cycle condition for detailed balance, is reproducible across African, Asian and European populations, and is strongest in CpG-linked cycles and CpG-transition edges, consistent with methylcytosine deamination. These results provide a thermodynamic decomposition of the human mutation field: most mutation bias is organized by a local equilibrium-like energy landscape aligned with genome composition, whereas the residual circulation highlights specific directional mutational mechanisms.
\end{abstract}

\maketitle

\section{Introduction}

DNA sequences evolve through stochastic moves on an enormous discrete space. A central question, common to statistical physics and evolutionary dynamics, is whether the observed directionality of such moves can be described by an effective scalar landscape, as in an equilibrium process satisfying detailed balance, or whether it contains irreducible circulation characteristic of non-equilibrium processes. In an equilibrium system, the logarithm of the ratio between a transition and its reverse coincides with the energy difference between the initial and final states involved. Equivalently, the sum of these log-ratios around any closed cycle vanishes. Nonzero cycle contributions therefore signal broken detailed balance and irreversible circulation \cite{Kolmogorov1936,Seifert2012,Esposito2010,Landi2021}.

Context-dependent mutation probabilities offer a rare opportunity to empirically test these ideas in an evolving molecular system. In human DNA, mutation probabilities depend strongly on the local sequence context, and seven-base context models explain much of the variability observed in population polymorphism data \cite{Aggarwala2016,Carlson2018,Liang2022}. These probabilities are usually treated as independent estimates of local mutation rates. However, they also define a graph of context-dependent substitutions: for each fixed six-base flanking sequence context, the graph has four nodes corresponding to the possible central bases, with directed edges representing substitutions between them. In particular, every substitution of the central base from $a$ to $b$ has a corresponding reverse substitution from $b$ to $a$. Since the associated transition probabilities need not coincide, the logarithm of their ratio defines an \textit{antisymmetric field} on the directed edges of this graph.

In this study, we indicate a seven-base context as $k_a$, where $k$ denotes the fixed six-base flanking context and $a\in\{A,C,G,T\}$ is the central nucleotide. A substitution $a\to b$ maps $k_a$ to $k_b$ while keeping the flank fixed. We assign to this directed edge the antisymmetric log-ratio
\begin{equation}
\phi_k(a,b) \equiv \phi(k_a\to k_b)=\log \frac{P(k_a\to k_b)}{P(k_b\to k_a)},\qquad a\ne b,
\label{eq:field}
\end{equation}
where $P(k_a\to k_b)$ is the context-dependent mutation probability. By construction, $\phi_k(a,b)=-\phi_k(b,a)$ and the mutation field is defined as the set of these values over all flanks and substitutions.
\ic{Unless otherwise specified, $P(k_a\to k_b)$ denotes the mutation probability averaged across the African, Asian, and European population-specific estimates reported by Aggarwala and Voight \cite{Aggarwala2016};  population-specific values were retained for
the cross-population and flux analyses. Data processing, population
averaging, and construction of the antisymmetric mutation field are
described in Methods, Section~\ref{sec:methods_data}.}

This construction is not specific to human mutations. Any context-dependent substitution table defines such a field. Its gradient component measures the part compatible with an effective energy landscape, whereas its curl component quantifies irreversible circulation around closed mutation cycles that cannot be represented by any scalar potential \cite{HodgeRanking2011}. The circulation around a mutation cycle, given by the sum of the forward-to-reverse log-ratios along the cycle, is known in stochastic thermodynamics as the cycle affinity \cite{schnakenberg1976network,rao2016nonequilibrium,liang2024thermodynamic}; a nonzero affinity signals nonequilibrium. The inferred energy should therefore be interpreted as an effective quantity rather than a direct measure of DNA stability.

The identification of reversible and irreversible components is closely connected to genome composition. Chargaff's first parity rule states that, in double-stranded DNA, adenine and thymine occur in nearly equal amounts, as do cytosine and guanine \cite{Zamenhof1950}. Chargaff's second parity rule extends this approximate symmetry to a single strand, where the frequency of a short word is often close to that of its reverse complement (from now on indicated as RC) \cite{Rudner1968I,Karkas1968,Rudner1968III,Baisnee2002,AlbrechtBuehler2006,Fariselli2021}. Recent work has also interpreted human genomic word frequencies through an effective thermodynamic potential, suggesting a pseudo-equilibrium organization of sequence composition \cite{FariselliMaritan2025}. What remains unclear is whether the directed mutation process itself is close to reversibility, and whether its irreversible component has recognizable biological content.
%Throughout the paper, the terms ``field'', ``reversible'', ``irreversible'', and ``energy'' are used in this operational graph-theoretic sense. 
%The field is an oriented quantity attached to edges of sequence space, not a spatial field in the nucleus. 
%Reversibility does not mean that individual mutations are physically undone, and the inferred energy is not a direct measurement of DNA stability. Rather, a reversible mutational field is one whose forward--reverse asymmetries can be written as differences of an effective mutational potential. An irreversible component is the residual circulation around closed mutation cycles that no scalar potential can represent.
From this perspective, CpG dinucleotides provide a natural mechanistic test case. Methylated cytosines are prone to deamination, leading to elevated C$\to$T and reverse-complement G$\to$A mutation probabilities \cite{DuncanMiller1980,CooperYoussoufian1988,FryxellMoon2005}. If the curl component of the mutation field were dominated by noise, high-affinity cycles would be broadly distributed across sequence contexts and nucleotide substitutions. Enrichment in CpG-linked states or CpG-transition edges would instead suggest that the residual circulation captures structured biochemical asymmetries.

\ic{The term equilibrium-like requires distinguishing stationarity from
detailed balance. Let $P(k_a)$ denote the reference-genome frequency
of a seven-base state and let $P^{(p)}(k_a\to k_b)$ denote the
\textit{population}-specific context-dependent substitution probability. For a
fixed flank $k$, approximate stationarity of the reference composition
requires the master-equation balance
% B_{k,b}^{(p)} sostituito con \partial_t P(k_b)
\begin{equation}
\begin{split}
\partial_t P(k_b) &= \sum_{a\neq b}
\Big[ P(k_a)P^{(p)}(k_a\to k_b) \\ &\qquad
- P(k_b)P^{(p)}(k_b\to k_a) \Big] \simeq 0
\end{split}
\label{eq:master_balance_intro}
\end{equation}
for each central state $b$. This condition states that the total
incoming and outgoing probability fluxes approximately balance.

Detailed balance is stronger. It requires each pairwise probability
current
\begin{equation}
J_{ab}^{(p)}(k) = P(k_a)P^{(p)}(k_a\to k_b)
- P(k_b)P^{(p)}(k_b\to k_a)
\label{eq:probability_current_intro}
\end{equation}
to vanish separately. Under detailed balance, the forward-to-reverse
log-ratio field is a pure gradient and every closed-cycle affinity
vanishes. A process may therefore be approximately stationary while
still sustaining weak irreversible currents around local mutation
cycles. Throughout this work, \emph{equilibrium-like} refers to this
combination: approximate master-equation balance, a dominant gradient
component, and a smaller but measurable circulation violating detailed
balance. The neural energy model probes the gradient structure, whereas
probability currents, entropy production, cycle affinities, and Hodge
decomposition quantify the residual irreversible component}.

\begin{figure*}[t]
    \centering
    \includegraphics[width=\textwidth]{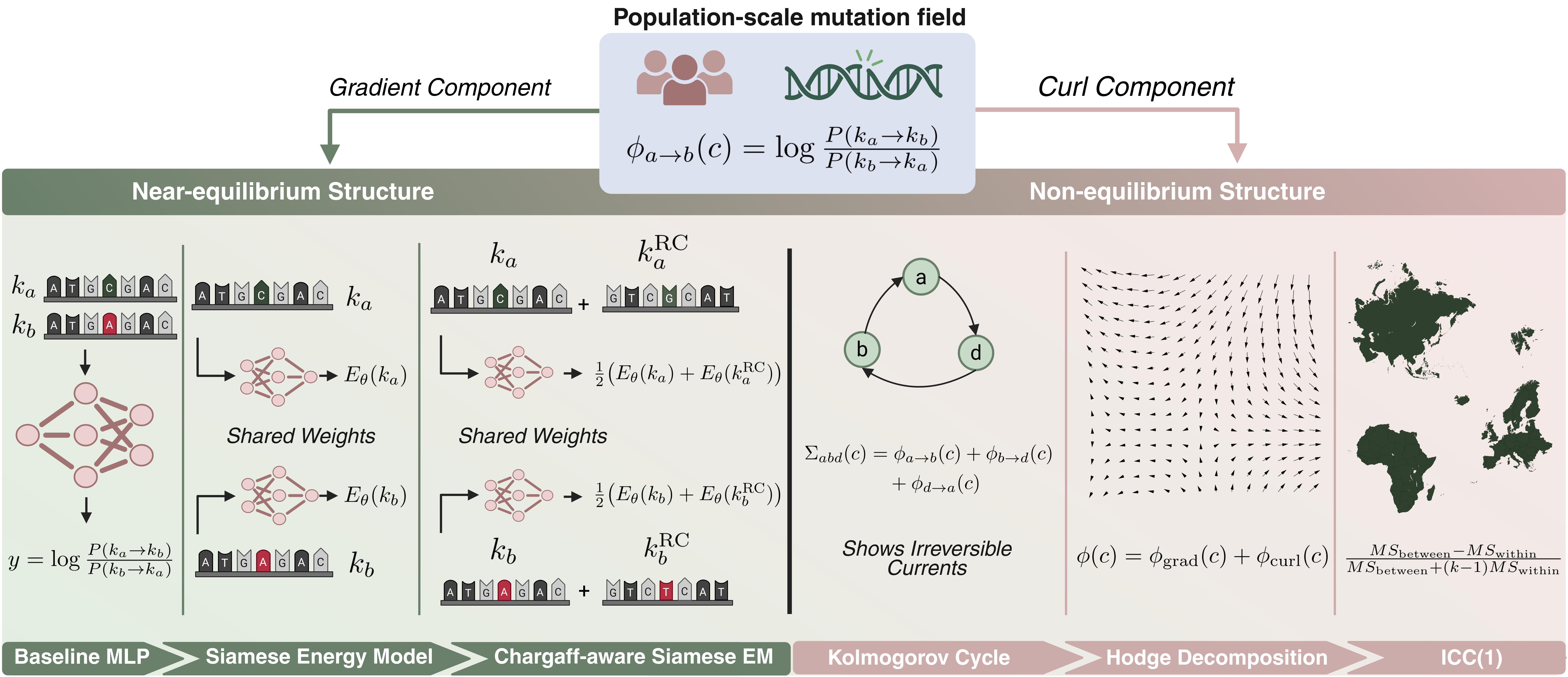}
    \caption{
    \textbf{Geometric decomposition of the human mutational field.}
    Population-scale strand-asymmetric mutation probabilities define a directed mutational field
    $\phi(k_a\to k_b)=\log[P(k_a\to k_b)/P(k_b\to k_a)]$ on local sequence space.
    The reversible, gradient-like component is represented by an energy learned with a Siamese neural architecture and saturates at a short interaction range ($r_{\mathrm{int}}\simeq3$--$4$ bases).
    Within-length aggregates of the learned energy define short-word composition proxies that recapitulate short-range genomic composition and approximate Chargaff symmetry.
    The irreversible, curl-like component is detected by non-zero Kolmogorov cycle affinities and isolated by Hodge decomposition; its high cross-population reproducibility indicates a structured non-equilibrium contribution rather than population-specific noise.
    }
    \label{fig:concept}
\end{figure*}

Here, we first test whether the human mutation field is compatible
with an approximately stationary and predominantly gradient-like
description. We then infer the effective energy landscape that best
explains its reversible component and assess whether the resulting
energies recover genomic composition and reverse-complement symmetry.
Finally, we quantify the residual irreversible component through
probability currents, entropy production, cycle affinities, and Hodge
decomposition, and investigate its biological association with
CpG-linked mutational mechanisms.  These steps are graphically summarized in Fig.~\ref{fig:concept}.

We characterize the biological origin of such irreversibility through CpG content and substitution types. Together, these results reveal that the human mutation field is organized by a dominant equilibrium-like structure, with a smaller but reproducible irreversible circulation enriched in CpG-associated mutational mechanisms.

\section{Results}

\subsection{The equilibrium-like structure of the human mutation field}

%Figure~\ref{fig:concept} summarizes the analysis. The reversible part of the field is the component that can be written as a difference of scalar energies. The irreversible part is the residual circulation, detected by non-zero sums around closed mutation cycles and isolated by Hodge projection. In this terminology, ``near-equilibrium'' means that the gradient component dominates prediction and Hodge projection; it does not require every local cycle to satisfy detailed balance exactly.

% \begin{figure*}[t]
%     \centering
%     \figbox[width=\textwidth]{figure1_prx_landscape_curl.pdf}
%     \caption{\textbf{Landscape--curl decomposition of the human mutational field.}
%     Context-dependent mutation probabilities define a directed field on local sequence space: for two seven-base contexts $k_a$ and $k_b$ that differ only at the central nucleotide, $\phi(k_a\to k_b)=\log[P(k_a\to k_b)/P(k_b\to k_a)]$.
%     If this field is reversible, the directed log-ratio can be written as the difference of a scalar landscape, $\phi(k_a\to k_b)=E(k_a)-E(k_b)$.
%     Broken detailed balance appears as non-zero affinity around closed mutation cycles, $\Sigma=\sum_{\mathrm{cycle}}\phi\neq0$.
%     Hodge decomposition separates the empirical field into a dominant gradient-like component and a residual curl-like component.
%     The gradient component provides an equilibrium-like background connected to genomic composition and reverse-complement symmetry, whereas the curl component highlights directional mutational processes, including CpG-associated methylcytosine deamination.}
%     \label{fig:pipeline}
% \end{figure*}

%\subsection{A local scalar landscape captures most mutation asymmetry}

 We first established a predictive reference without imposing reversibility or an equilibrium-like structure. An unconstrained multilayer perceptron receives the ordered pair $(k_a,k_b)$ as input and is trained directly to predict $\phi_k(a,b)$ \ic{(Methods, Section~\ref{sec:methods_mlp})}. Because the model sees source and target sequences jointly, it can in principle learn any antisymmetric function of the pair, including features not representable by a scalar potential. Across 20 independent train--test splits, this model predicts log-ratios for mutations excluded from training with Pearson correlation $r = 0.948 \pm 0.002$, and mean squared error equal to $0.143 \pm 0.005$. The uncertainties denote standard deviations across splits. This value for $r$ constitutes an empirical predictive reference when a fixed thermodynamic structure is not imposed a priori.

\begin{figure*}[t]
    \centering
    \includegraphics[width=\textwidth]{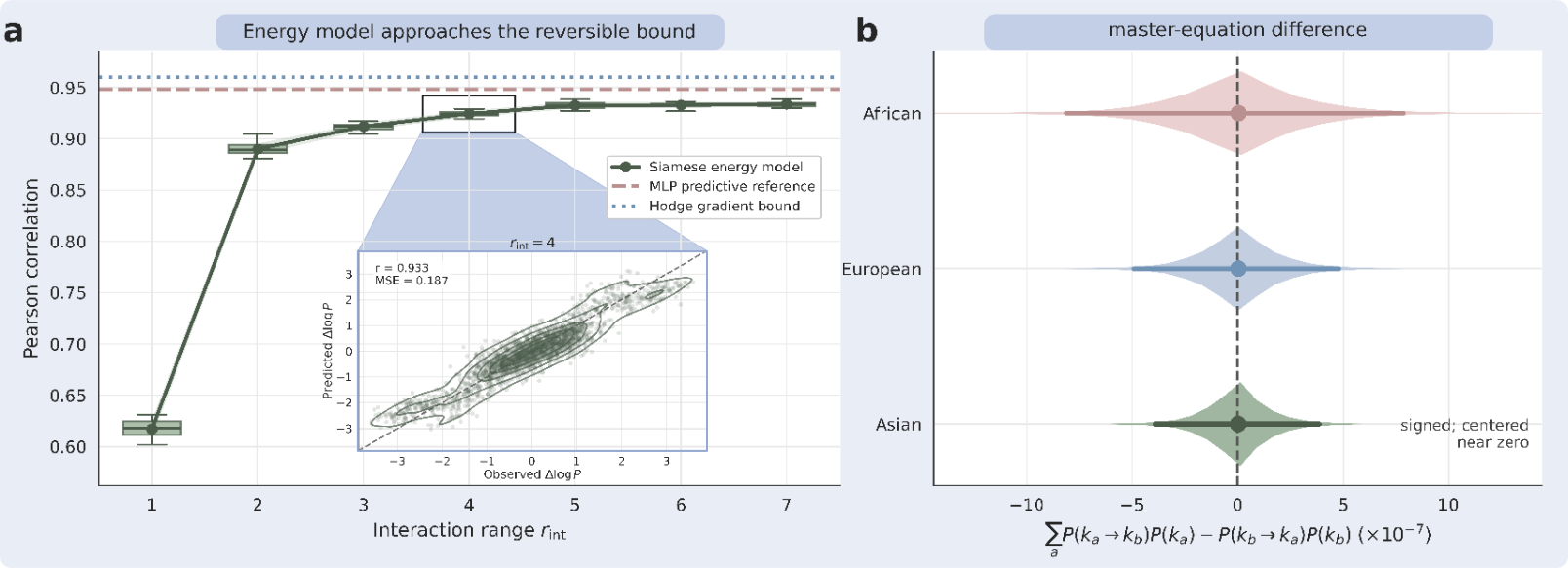}
    \caption{
    \textbf{A local energy model captures the dominant reversible
    component of an approximately stationary human mutation field.}
    \textbf{a}, Cross-validation performance of the Siamese energy
    model as a function of the interaction range
    $r_{\mathrm{int}}$. Pearson correlation increases sharply between
    one and two bases, improves more gradually up to approximately
    three to four bases, and then approaches a plateau. The horizontal
    dashed lines indicate the predictive reference provided by the
    unconstrained multilayer perceptron and the empirical reversible
    ceiling obtained from the Hodge gradient projection. Box plots
    summarize performance across independent data splits.The inset shows a representative predicted-versus-observed comparison
on held-out mutations for $r_{\mathrm{int}}=4$, the shortest
interaction range close to the performance plateau; points represent
held-out transitions, density contours highlight empirical point
density, and the dashed diagonal indicates perfect agreement.
Detailed Pearson and MSE values are reported in Table~\ref{tab:siamese_performance_by_k}.
    \textbf{b}, Signed master-equation balance terms obtained by
    combining population-specific mutation probabilities with
    reference-genome seven-base frequencies. Distributions are centered
    near zero for the African, European, and Asian estimates, supporting
    approximate stationarity of the reference composition, while their
    finite width reflects context-specific deviations from exact
    balance. Values are scaled by $10^{-7}$.}
    \label{fig:energy}
\end{figure*}

To test the reversible hypothesis directly, we trained a Siamese energy model. A shared neural network maps each seven-base sequence $s$ to a scalar $E_{\theta}(s)$, and the predicted log-ratio for a mutation is constrained to be
\begin{equation}
\hat{\phi}_k(a,b)=E_{\theta}(k_a)-E_{\theta}(k_b).
\label{eq:energy_model}
\end{equation}
The model, \ic{described in Methods, Section~\ref{sec:methods_siamese}}, can therefore succeed only by learning an energy landscape whose pairwise differences reproduce the directional mutation asymmetries. We varied the first convolutional kernel size $r_{\mathrm{int}}$ from one to seven bases to estimate the effective interaction range. Performance rose sharply from $r\simeq0.62$ at $r_{\mathrm{int}}=1$ to $r\simeq0.89$ at $r_{\mathrm{int}}=2$, then approached a plateau of $r\simeq0.92$--0.93 for $r_{\mathrm{int}}=3$--4 and beyond. Increasing the window beyond four bases yielded only marginal improvement, \ic{as shown in Fig.~\ref{fig:energy}a.}  Thus, most reversible structure in the measured mutation field is captured by short-range interactions.

The constrained energy model approaches, but does not exceed, the unconstrained predictive reference. The gap is not only an architectural comparison; it measures the cost of imposing a gradient constraint on a field that may contain irreversible components. The high value of the resulting correlation in the Siamese network nevertheless shows that the leading component of human mutation directionality is well described by an equilibrium description in terms of an effective energy landscape.

\ic{To assess approximate stationarity directly, we combined the
population-specific mutation probabilities with reference-genome
seven-base frequencies and evaluated the signed master-equation
balance terms defined in Eq.~\ref{eq:master_balance_intro}.
Figure~\ref{fig:energy}b shows that these distributions are
centered near zero in the African, Asian, and European estimates,
although their finite width reveals context-specific deviations from
exact balance. Thus, the reference-genome composition is approximately,
but not exactly, stationary under the measured mutation probabilities.}

\subsection{Energy landscape substantially captures genome composition and reverse-complement symmetry}

A key test to uncover the information content of the equilibrium-like structure is whether an energy inferred solely from mutation probabilities contains information about genome composition. The model is not trained on genomic word frequencies and is not constrained to obey Chargaff symmetry. We therefore analyzed to what extent short-word summaries of the learned energy landscape recapitulate empirical word composition.

\begin{figure*}[t]
    \centering
    \includegraphics[width=\textwidth]{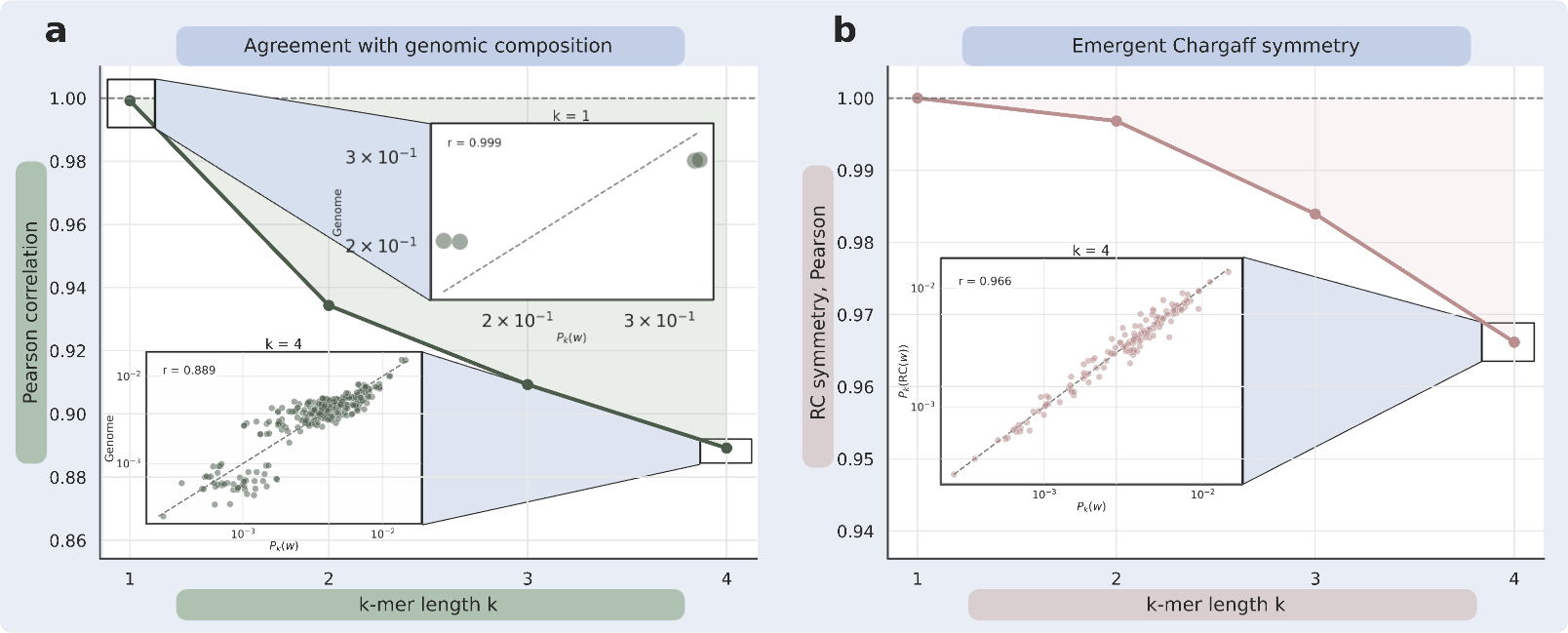}
    \caption{
    \textbf{Mutation-trained energies recapitulate genomic composition
    and reverse-complement symmetry.}
    The within-length composition proxies
    $P_{\theta}^{(\ell)}(w)$ were derived exclusively from mutation
    log-ratios, without training on genomic word frequencies.
    \textbf{a}, Pearson correlation between the model-derived
    composition proxies and observed GRCh38/hg38 word frequencies for
    word lengths $\ell=1,\ldots,4$. Agreement remains strong but
    gradually decreases as word length increases. The insets show the
    direct word-level comparisons for $\ell=1$ and $\ell=4$,
    corresponding respectively to the shortest and most demanding word
    lengths considered. Both inset axes are logarithmic, and dashed
    diagonals indicate equality.
    \textbf{b}, Reverse-complement symmetry of the model-derived
    composition proxies, measured as the Pearson correlation between
    $P_{\theta}^{(\ell)}(w)$ and
    $P_{\theta}^{(\ell)}(\mathrm{RC}(w))$. The inset shows the
    word-level comparison for $\ell=4$ on logarithmic axes, with the
    dashed diagonal indicating equality. Together, the two panels show
    that the energy landscape inferred solely from mutation
    directionality contains substantial information about short-range
    genomic composition and approximately preserves Chargaff symmetry.
    }
    \label{fig:composition}
\end{figure*}

For a word $w$ of length $\ell\le4$, let ${\cal C}(w)$ be the set of seven-base contexts in which $w$ occurs and contains the mutated position, i.e., the central nucleotide. We define the aggregate score
\begin{equation}
E_{\mathrm{agg}}(w)=-\log\sum_{s\in{\cal C}(w)}\exp[-E_{\theta}(s)]
\label{eq:eagg}
\end{equation}
that resembles a free energy associated with the set $C(w)$. Therefore, the Boltzmann weight associated with a word $w$ within the ensemble of words of length $\ell$ is
\begin{equation}
P_{\theta}^{(\ell)}(w)=\frac{\exp[-E_{\mathrm{agg}}(w)]}{\sum_{|u|=\ell}\exp[-E_{\mathrm{agg}}(u)]} \;.
\label{eq:proxy}
\end{equation}
This quantity defines a composition proxy for words of fixed length. These proxies strongly agree with empirical genomic word frequencies computed from GRCh38/hg38 (Fig.~\ref{fig:composition}). The Pearson correlation is approximately $0.999$ for words of length
one and monotonically decreases to $0.889$ for words of length four
(Fig.~\ref{fig:composition}a). The corresponding word-level
comparisons are shown in the $\ell=1$ and $\ell=4$ insets. The same proxies approximately satisfy Chargaff's second parity rule (see Fig.~\ref{fig:composition}b),
\begin{equation}
P_{\theta}^{(\ell)}(w)\simeq P_{\theta}^{(\ell)}(\RC(w)),
\end{equation}
for short words. Figure~\ref{fig:composition} separates agreement with empirical genomic
composition from reverse-complement symmetry, while the insets show the
word-level relationships underlying the summary correlations.
Thus, an energy landscape learned solely from mutation directionality recapitulates substantial information about the compositional organization and reverse-complement symmetry of the human genome.

\begin{figure*}[t]
    \centering
    \includegraphics[width=\textwidth]{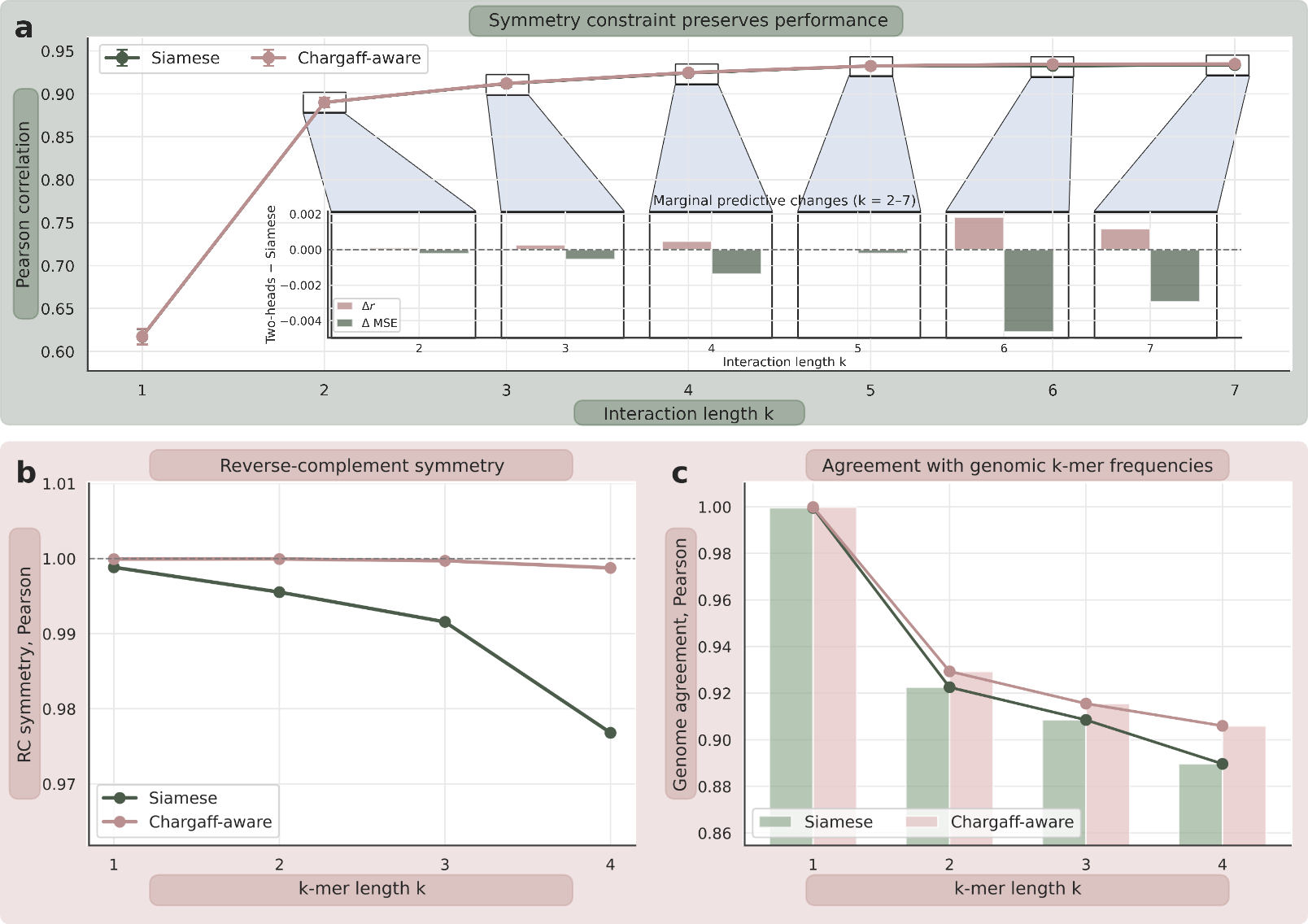}
    \caption{
    \textbf{A reverse-complement-aware architecture preserves mutation
    prediction while enforcing Chargaff symmetry.}
    \textbf{a}, Cross-validation Pearson correlation of the standard
    Siamese energy model and the Chargaff-aware two-head architecture
    across interaction ranges. The two performance curves are nearly
    superimposed. The magnified lower panels report the corresponding
    marginal changes in Pearson correlation,
    $\Delta r=r_{\mathrm{two\mbox{-}head}}-r_{\mathrm{Siamese}}$, and
    mean squared error,
    $\Delta\mathrm{MSE}
    =\mathrm{MSE}_{\mathrm{two\mbox{-}head}}
    -\mathrm{MSE}_{\mathrm{Siamese}}$,
    for $r_{\mathrm{int}}=2,\ldots,7$, showing that the predictive
    differences between the two architectures are small.
    \textbf{b}, Reverse-complement symmetry of the derived composition
    proxies as a function of word length. The explicit symmetry
    constraint maintains correlations close to one, particularly at
    longer word lengths.
    \textbf{c}, Agreement between the model-derived composition proxies
    and observed genomic word frequencies. The Chargaff-aware model
    preserves, and slightly stabilizes, agreement with genomic
    composition while enforcing reverse-complement consistency.
    Together, these results show that Chargaff symmetry can be
    incorporated as an architectural constraint without sacrificing
    the ability to model context-dependent mutation probabilities.
    }
    \label{fig:chargaff}
\end{figure*}

We also trained a Chargaff-aware two-head model in which each input sequence is paired with its reverse complement \ic{(Methods, Section~\ref{sec:methods_chargaff})}. The main prediction uses the average of the two energies for each reverse-complement pair, and an auxiliary loss penalizes $E_{\theta}(s)-E_{\theta}(\RC(s))$. This explicit symmetry constraint changes predictive accuracy only
marginally, with differences in Pearson correlation of order $10^{-4}$--$10^{-3}$ and small changes in mean squared error across the tested interaction ranges. Its main effect is to stabilize the inferred composition proxies, preserving near-perfect reverse-complement correlations at larger word lengths (Fig.~\ref{fig:chargaff}). The result suggests that the equilibrium-like energy model has already learned most of the signal associated with Chargaff symmetry; the remaining mismatch to the unconstrained predictive baseline is therefore likely to originate from a genuinely irreversible, nonequilibrium component.

\subsection{\ic{Probability currents and cycle affinities reveal a nonequilibrium contribution conserved across populations}}

\begin{figure*}[t]
    \centering
    \includegraphics[width=\textwidth]{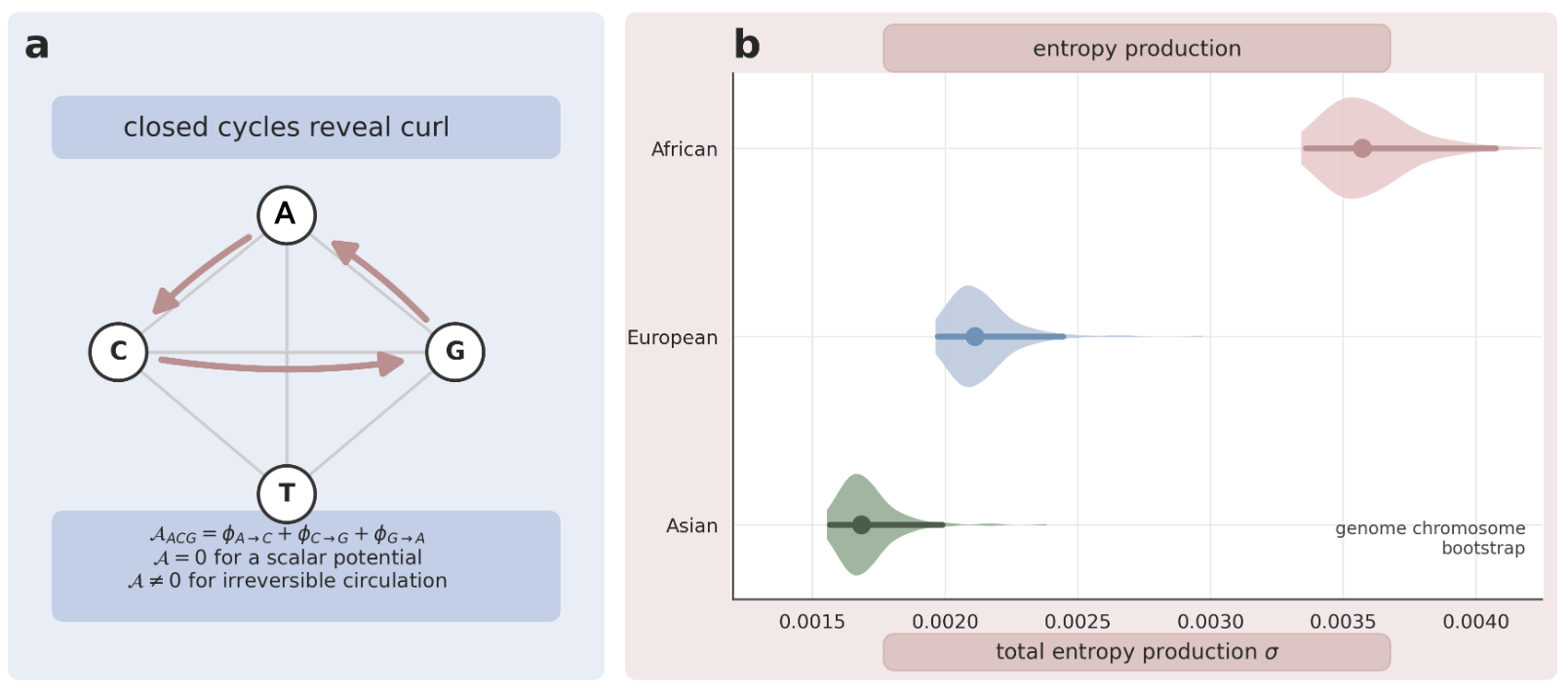}
    \caption{
    \textbf{Closed-cycle affinities and entropy production reveal a
    residual irreversible component of the human mutation field.}
    \textbf{a}, Schematic of the four-state central-nucleotide mutation
    graph for a fixed flanking context. The highlighted cycle
    $A\to C\to G\to A$ has affinity
    $\mathcal{A}_{ACG}
    =\phi_{A\to C}+\phi_{C\to G}+\phi_{G\to A}$.
    For any field generated by a scalar potential, the contributions
    cancel around every closed loop and
    $\mathcal{A}=0$. A nonzero cycle affinity therefore identifies
    irreversible circulation that cannot be represented by an
    energy-only model.
    \textbf{b}, Chromosome-bootstrap distributions of the total entropy
    production $\sigma^{(p)}$, evaluated by combining
    population-specific mutation probabilities with reference-genome
    seven-base composition. The entropy production is positive but
    small in all three populations, indicating weak yet systematic
    irreversibility relative to the observed genomic composition. The
    moderately larger African estimate should be interpreted cautiously
    in view of the greater variability of the corresponding
    population-derived mutation-probability estimates.
    }
    \label{fig:irreversibility}
\end{figure*}

\ic{Approximate master-equation balance does not require every pairwise
probability current defined in Eq.~\ref{eq:probability_current_intro}
to vanish. Stationarity constrains the sum of the currents entering
and leaving each state, whereas detailed balance requires
$J_{ab}^{(p)}(k) = 0$ separately for every edge. Consequently, nonzero
currents may circulate around closed mutation loops even when the net
master-equation balance is close to zero.

The irreversibility associated with these currents for each population $p$ can be summarized through the \textit{entropy production} ($\sigma^{(p)}$) evaluated at the reference-genome composition,
\begin{equation}
\sigma^{(p)} = \sum_k \sum_{a<b} J_{ab}^{(p)}(k) \log \frac{
P(k_a)P^{(p)}(k_a\to k_b)
}{P(k_b)P^{(p)}(k_b\to k_a)}
\label{eq:entropy_production}
\end{equation}
The entropy production vanishes under pairwise detailed balance and is
non-negative, becoming positive whenever at least one pair of forward
and reverse fluxes is imbalanced. Because $P(k_a)$ is taken from the reference genome rather
than inferred as the exact stationary distribution of each
population-specific transition matrix, $\sigma^{(p)}$ should be
interpreted as entropy production evaluated relative to the observed
genomic composition, rather than as an absolute physical rate per unit
time. Chromosome-level bootstrap resampling was used to propagate uncertainty
in the reference-genome composition. At each bootstrap iteration, chromosomes were sampled with replacement,
seven-base context frequencies were recomputed, and
$\sigma^{(p)}$ was recalculated separately for each population.

Across populations, the resulting entropy-production distributions
were positive but small (Fig.~\ref{fig:irreversibility}b), consistent
with weak but systematic irreversibility relative to the reference
composition. Their proximity to zero indicates that this departure is
limited in magnitude, although consistently detectable across
bootstrap replicates. The moderately larger value observed for the African
population should be interpreted cautiously, because this population
also carries greater genetic variability and may yield more variable
mutation-probability estimates.

A complementary, composition-independent signature of irreversibility
is provided by closed-cycle affinities. Under detailed balance, the
affinity of every closed mutation cycle vanishes; nonzero affinities
therefore identify a component of the mutation field that cannot be
represented by a scalar potential.}

For each fixed flank $k$ and each ordered three-base cycle of the central nucleotide, $a \to b \to d \to a$, we computed the corresponding cycle affinity \cite{schnakenberg1976network} (see Fig.~\ref{fig:irreversibility}a):
%\dmb{While} neural \dmb{networks} test whether an energy landscape can approximate the field, we here address a stricter geometric question: which part of the empirical field is inaccessible to any scalar potential? A purely gradient field has zero affinity around every closed cycle.
\begin{equation}
\mathcal{A}_k(a,b,d)=\phi_k(a,b)+\phi_k(b,d)+\phi_k(d,a) \;.
\label{eq:cycle}
\end{equation}
The six flanking positions give $4^6$ possible flanks. There are four choices of three central bases, each defining a unique mutation cycle up to orientation. We used one arbitrarily chosen orientation for each triplet, giving $4^6\times4=16{,}384$ cycle affinities. The reverse orientation has the opposite sign and was not counted separately.
Because reversing the orientation of a closed cycle changes $\mathcal{A}$ to $-\mathcal{A}$, the mean affinity is expected to vanish by symmetry and is therefore not evidence for detailed balance. We instead consider the spread of these cycle affinities and their orientation-independent magnitude $|\mathcal{A}|$. In particular, we found a substantial standard deviation around zero, $sd_{\mathcal{A}}=0.773$, and a median absolute affinity $m_{|\mathcal{A}|} = 0.451$, with probability $P(|\mathcal{A}|>0.5) = 0.462$ of observing an orientation-independent affinity exceeding $0.5$. Therefore, substantial nonzero affinities are broadly distributed across local mutation cycles and signal the presence of an irreversible component that cannot be explained by an effective equilibrium model. The cycle affinities are also highly reproducible across populations. Using population-specific mutation probabilities for African, Asian, and European samples, a one-way random-effects intraclass correlation gives $\mathrm{ICC}=0.91$ for the cycle affinities. Most context-to-context variation in the nonequilibrium component is therefore shared across populations rather than being population-specific noise. \ic{The intraclass-correlation estimator and the definition of the
cycle-flank items used in this analysis are given in
Methods, Section~\ref{sec:methods_cycles}.}

Because the gradient part of the mutation field has zero
circulation around every closed cycle, the cycle affinity depends only
on the curl component. To quantify the relative contribution of these
two components, we performed a Hodge projection of the mutation graph defined by the four possible central
nucleotides for each fixed flank. The graph contains 12 directed
edges corresponding to all possible substitutions. \ic{Let
$B\in\mathbb{R}^{12\times4}$ denote its edge--node incidence matrix,
with $(Bg)_{a\to b}=g(a)-g(b)$. The best-fitting gradient field was
obtained by least squares,
\begin{equation}
g_k^\ast
=
\arg\min_{g\in\mathbb{R}^4}
\left\|\phi_k-Bg\right\|_2^2,
\label{eq:hodge_projection}
\end{equation}
with one arbitrary gauge fixed to remove the additive freedom of the
scalar potential. The field was then decomposed
into the best-fitting gradient component and an orthogonal residual
cycle component:
\begin{equation}
\phi_k
=
\phi_k^{\rm grad}
+
\phi_k^{\rm curl},
\qquad
\phi_k^{\rm grad}=Bg_k^\ast.
\label{eq:hodge_components}
\end{equation}}
The curl fraction for each context quantifies the relative weight
of the non-gradient component and is defined as
\begin{equation}
f_{\rm curl}(k) =
\frac{\left\|\phi_k^{\rm curl}\right\|_2^2
}{\left\|\phi_k^{\rm grad}\right\|_2^2
+\left\|\phi_k^{\rm curl}\right\|_2^2}.
\label{eq:curl_fraction}
\end{equation}
The mean curl fraction over all contexts is $0.11$, and the median is
$0.07$. The mutation field is therefore mostly reversible, but retains
a measurable non-gradient irreversible component that reflects nonzero cycle affinities.

The Hodge projection also defines a data-derived ceiling for
energy-only models. \ic{We define this empirical reversible ceiling as
\begin{equation}
r_{\max}
=
\operatorname{corr}
\left(
\bigoplus_k \phi_k,
\bigoplus_k \phi_k^{\rm grad}
\right),
\label{eq:rmax}
\end{equation}
where $\bigoplus_k$ denotes concatenation across all fixed flanks
and directed substitutions.} The Pearson correlation between the
mutation field estimated from the data and its best gradient
approximation is $r_{\max}\simeq0.96$. This is a geometric bound
computed from the empirical field, distinct from the predictive
reference provided by the unconstrained neural network on
mutations excluded from training. The fact that the Siamese energy
model approaches this ceiling indicates, once again, that its residual
prediction error largely reflects the irreversible component of the
mutation field rather than limitations of the model architecture.

\subsection{Irreversible circulation is enriched in CpG-associated mechanisms}

\begin{figure*}[t]
    \centering
    \includegraphics[width=\textwidth]{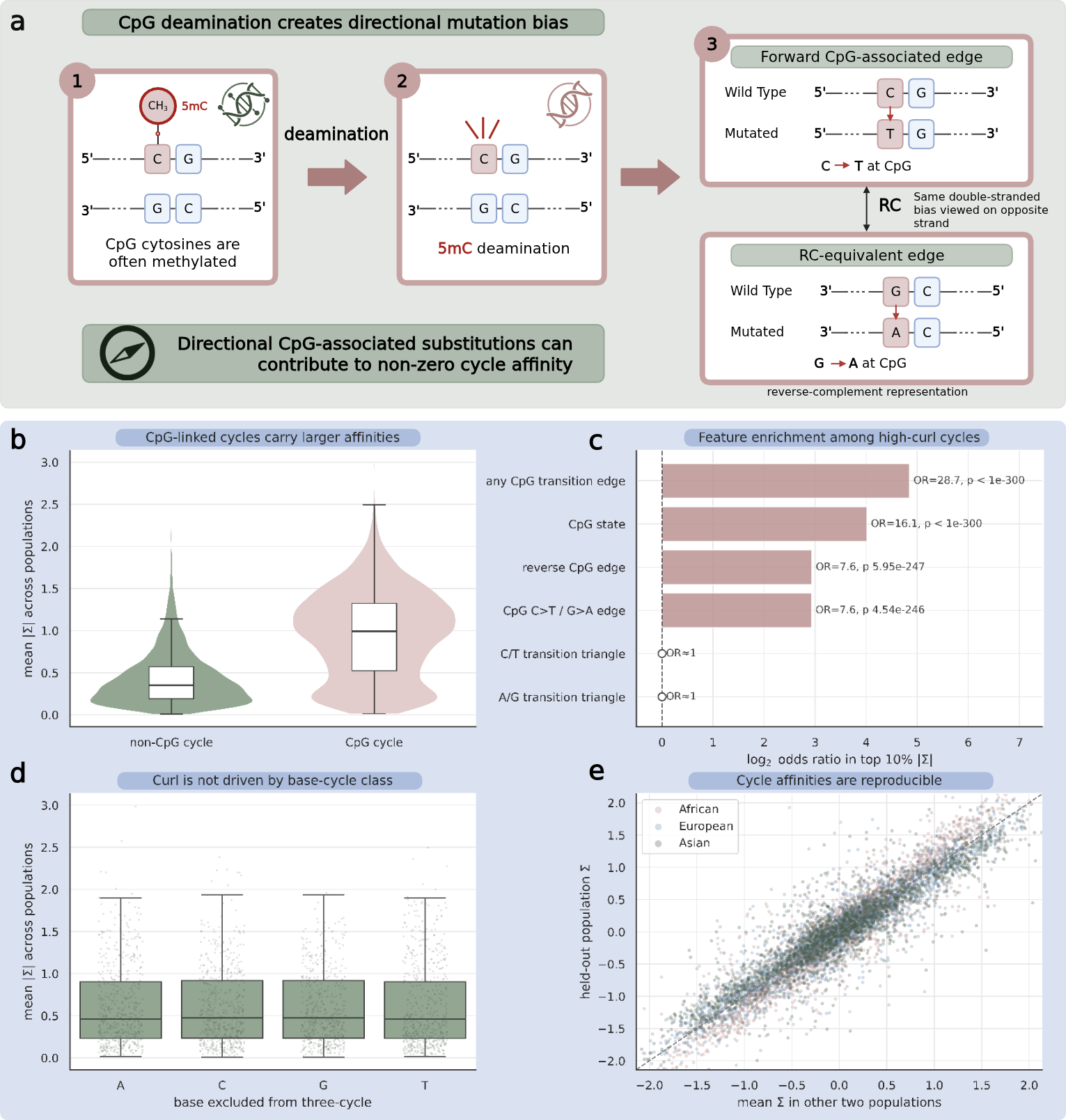}
\caption{
    \textbf{The strongest irreversible cycles are CpG-linked and
    reproducible across populations.}
    \textbf{a}, CpG-associated mutational mechanism. Deamination of
    5-methylcytosine produces a $C\to T$ substitution at a CpG site,
    represented on the opposite strand as the
    reverse-complement-equivalent $G\to A$ substitution. Directional
    imbalance between these mutations and their reverse directions can
    contribute to nonzero cycle affinities.
    \textbf{b}, Distribution of mean absolute cycle affinity
    $|\mathcal{A}|$ for CpG-linked and non-CpG cycles.
    \textbf{c}, Enrichment of biological annotations among high-curl
    cycles, defined as the top 10\% by $|\mathcal{A}|$. Bars show
    $\log_2$ odds ratios; the dashed line indicates no enrichment.
    CpG-state and CpG-transition annotations are strongly enriched,
    whereas generic C/T and A/G transition-triangle controls remain
    close to the null expectation.
    \textbf{d}, Negative-control analysis grouping cycles by the
    nucleotide excluded from the three-base cycle.
    \textbf{e}, Leave-one-population-out reproducibility of signed cycle
    affinities. Together, these analyses show that the irreversible
    component is biologically structured, concentrated in CpG-linked
    contexts, and shared across populations.
    }
    \label{fig:cpg}
\end{figure*}

We next asked whether the curl component was biologically structured.
\ic{CpG dinucleotides provide a natural mechanistic candidate because
cytosines in CpG contexts are frequently methylated at carbon 5, and
spontaneous deamination of 5-methylcytosine produces thymine
\cite{DuncanMiller1980,CooperYoussoufian1988,FryxellMoon2005}.
This process generates an elevated $C\to T$ transition when the
mutable central cytosine is followed by G. When the same
double-stranded event is represented from the opposite strand, the
corresponding reverse-complement substitution is $G\to A$, with the
mutable central guanine preceded by C. Because these directional
CpG-associated transitions can generate imbalances between forward and
reverse mutation edges, they are expected to contribute to nonzero
cycle affinities. The strand-oriented representation of these CpG-associated
substitutions is illustrated in Fig.~\ref{fig:cpg}a. The schematic emphasizes that the $C\to T$ and $G\to A$
representations correspond to the same double-stranded mutational bias
viewed from opposite strand orientations.
A CpG site is defined as a cytosine followed immediately by guanine on
the same DNA strand, with the ``p'' denoting the intervening phosphate
bond.}

For each three-base cycle, we annotated whether at least one
state contained a CpG dinucleotide overlapping the mutable central
position and whether the cycle contained a CpG-associated transition
edge \ic{(Methods, Section~\ref{sec:methods_cpg})}. Specifically, a
CpG-associated forward edge was defined as $C\to T$ when the central
cytosine was followed by G, or as the reverse-complement-equivalent
$G\to A$ substitution when the central guanine was preceded by C. The
opposite $T\to C$ and $A\to G$ directions were tracked separately. We
then compared these annotations with the mean absolute cycle affinity
across populations.

\ic{The categories in Fig.~\ref{fig:cpg}c distinguish CpG-specific
annotations from broader mutation-class controls. A \emph{CpG state}
indicates that at least one state in the cycle contains a CpG
overlapping the mutable central position. A \emph{CpG-associated
forward edge} denotes a $C\to T$ substitution at a central CpG
cytosine or its reverse-complement-equivalent $G\to A$ substitution;
the opposite $T\to C$ and $A\to G$ directions are classified as
\emph{reverse CpG edges}. \emph{Any CpG-transition edge} is the union
of these forward and reverse categories. The \emph{C/T} and
\emph{A/G transition triangles} instead serve as generic controls,
indicating that the three-node cycle contains the corresponding
transition pair regardless of CpG context.}

CpG-linked cycles carried substantially larger affinities than
non-CpG cycles (Fig.~\ref{fig:cpg}b). \ic{The shift involves the broader
distribution rather than only a small number of extreme cycles,
indicating a systematic association between CpG context and
irreversible circulation.} Among the top 10\% of cycles ranked by
$|\mathcal{A}|$, CpG-related features were strongly enriched: cycles
containing any CpG-transition edge showed an odds ratio of
approximately 28.7, while cycles containing a CpG state showed an odds
ratio of approximately 16.1 (Fig.~\ref{fig:cpg}c). \ic{The enrichment is
therefore strongest when the cycle explicitly traverses a
CpG-associated mutation edge.}
We emphasize the magnitude of these effect sizes
because nearby cycles may share contexts or edges and are therefore
not fully statistically independent.

Generic base-cycle composition did not explain the effect. Grouping
cycles according to the nucleotide excluded from the three-base cycle
produced nearly indistinguishable $|\mathcal{A}|$ distributions
(Fig.~\ref{fig:cpg}d), providing a negative control against the
possibility that high curl simply reflects broad
transition--transversion composition or a particular three-base cycle
class.

The signed cycle affinities were also highly reproducible across
populations. In leave-one-population-out analyses, the affinity
measured in the held-out population was well predicted by the mean
affinity in the other two populations (Fig.~\ref{fig:cpg}e). \ic{The concentration of points around the equality line indicates that
both the sign and the relative magnitude of the strongest cycle
affinities are largely shared across populations.} Thus, the
irreversible component is not only geometrically measurable, but is
also concentrated in biologically interpretable CpG-linked contexts
and shared across populations. The highest-affinity individual cycles
are listed in Appendix~\ref{app:additional}, \ic{where CpG-linked cycles visibly dominate
the top-ranked examples.}

CpG-linked cycles nevertheless do not exhaust the irreversible
component. Non-CpG cycles also display nonzero affinities, indicating
that the curl captures a broader set of context-dependent directional
asymmetries beyond CpG hypermutability alone.

\section{Discussion}

The main result of this study is that human context-dependent mutation probabilities define an empirical non-equilibrium field with a dominant landscape component and a reproducible residual circulation. Most forward--reverse asymmetries behave as if local DNA contexts occupy an effective mutational landscape, and this landscape is strongly aligned with short-word genome composition and reverse-complement symmetry. The remaining non-gradient component is smaller, but enriched in CpG-associated transitions, indicating that it captures specific directional mutational processes rather than random noise.

In this sense, the work extends a stationary thermodynamic description of genome composition to a geometric description of directed mutation probabilities on local sequence space. The dominant part of the field is captured by a scalar potential landscape: a neural model constrained to predict only energy differences explains held-out mutation log-ratios with correlation about 0.93 and approaches the best reversible projection of the measured field. Thus, the pseudo-equilibrium interpretation of genome composition is visible not only in stationary word frequencies, but also in the directionality of mutation probabilities.

The effective interaction range is short. Most predictive gain is obtained by extending the model to three or four bases, after which performance saturates. This does not imply that longer genomic features are irrelevant to mutation in general; replication timing, chromatin state, methylation, transcription and repair can all shape mutation probabilities. Rather, within the seven-base context data analyzed here, the reversible component of forward--reverse asymmetry is largely local.

This local landscape is not only predictive: it also carries compositional information. Aggregated short-word scores correlate strongly with empirical genomic frequencies and recapitulate reverse-complement symmetry. Because the model is not trained on genome composition, this agreement provides an independent link between mutation directionality, effective energies and Chargaff-like organization. These aggregate scores should be interpreted as normalized short-word proxies, rather than as a direct reconstruction of a full seven-base stationary distribution.

The same decomposition also identifies what the energy landscape cannot explain. Non-zero cycle affinities, a non-zero Hodge curl fraction and a reproducible cross-population pattern all point to a low-amplitude but broadly distributed structured non-equilibrium component. The field is therefore near-equilibrium in a precise sense: most of the predictive and Hodge-projected signal is gradient-like, but the stricter Kolmogorov cycle condition is measurably violated. Geometrically, the empirical mutation log-ratio field contains circulation that no scalar potential can represent. Identifying the molecular origin of this circulation will require data stratified by replication timing, chromatin state, methylation, transcriptional asymmetry, repair pathways and cell type. The present analysis does not assign the curl to any one of these mechanisms.

Having defined this residual geometrically, the CpG enrichment analysis asks whether it has biological content. The largest irreversible affinities are not uniformly distributed across sequence space, but are concentrated in CpG-linked states and CpG-transition edges, whereas generic base-cycle classes do not show comparable enrichment. This pattern is consistent with the known directionality of methylcytosine deamination and supports the view that the curl captures structured mutational asymmetries rather than a purely geometric artifact. At the same time, CpG hypermutability accounts for a prominent and biologically interpretable part of a broader irreversible structure.

The probabilities used here are inferred from population polymorphism data. They therefore reflect mutation processes as observed through segregating variants and may include effects of ascertainment, local selection, demographic history and estimation noise. In particular, GC-biased gene conversion could contribute to the non-CpG residual curl in polymorphism-derived estimates, because it can favor the transmission or fixation of G/C over A/T alleles in recombination-associated regions without representing a primary mutation-rate asymmetry \cite{DuretGaltier2009}. The high cross-population reproducibility of the cycle signal argues against a purely population-specific artifact.

Practically, this decomposition provides both a compact null model for context-dependent mutation and a screening tool for mechanism-specific asymmetries. It separates the dominant equilibrium-like background from directional residuals, links mutation bias to short-range genome composition without fitting genomic frequencies directly, and can be applied to de novo mutations, cancer mutational signatures, species comparisons, or data stratified by methylation, replication timing, chromatin state and repair status.

In summary, the proposed framework turns context-dependent mutation tables into interpretable maps: an equilibrium-like mutational background aligned with genome composition, and a residual component highlighting directional processes that cannot be explained by that background. In humans, this residual map recovers CpG methylation-associated mutability as a dominant signal and reveals additional non-CpG asymmetries that merit mechanistic investigation. This landscape--curl decomposition may therefore provide a general tool for comparing mutational mechanisms across populations, tissues, cancers and species.

\section{Methods}
\label{sec:methods}

The analysis consisted of five steps: construction of the antisymmetric mutation log-ratio field; unconstrained prediction as an empirical predictive reference; energy-constrained Siamese prediction of the reversible component; derivation of short-word composition proxies; and quantification of irreversible circulation by cycle affinities, Hodge projection, cross-population reproducibility and CpG annotation.

\subsection{Data sources and notation}
\label{sec:methods_data}

We analyzed human germline substitution probabilities in seven-base contexts from population-scale estimates for African, Asian and European samples reported by Aggarwala and Voight \cite{Aggarwala2016}. Unless otherwise stated, the mutation probability used in model training was the arithmetic mean across the three populations,
\begin{equation}
P(k_a\to k_b)=\frac{1}{3}\sum_{p\in\{\mathrm{Afr},\mathrm{Asn},\mathrm{Eur}\}}P^{(p)}(k_a\to k_b).
\end{equation}
Population-specific probabilities were retained for
master-equation, entropy-production, cycle-affinity, and
cross-population reproducibility analyses. Genomic word frequencies for word lengths $\ell=1,\ldots,7$ were computed from GRCh38/hg38 chromosome sequences. The notation $k_a$ denotes a seven-base context with fixed six-base flank $k$ and central base $a$, $r_{\mathrm{int}}$ denotes neural interaction range, and $\theta$ denotes trainable neural-network parameters. \\
\ic{For each fixed flank $k$ and each pair of distinct central bases
$a,b\in\{A,C,G,T\}$, the mutation field was constructed using the
antisymmetric log-ratio defined in Eq.~\ref{eq:field}. Only
substitutions with valid probability estimates in both the forward and
reverse directions were retained. Whenever both directions were
available, the two oriented examples
\[
(k_a,k_b,\phi_k(a,b))
\qquad\text{and}\qquad
(k_b,k_a,-\phi_k(a,b))
\]
were included in the data set, ensuring exact antisymmetry of the
training targets.}

\ic{\subsection{Neural Network Approaches}}
\label{sec:methods_nn}

\subsubsection{Unconstrained neural baseline}
\label{sec:methods_mlp}

Each seven-base context was one-hot encoded as a $7\times4$ array and flattened to 28 inputs. The baseline multilayer perceptron received the concatenated encodings of $(k_a,k_b)$, giving a 56-dimensional input. The network had three hidden layers with 8, 16 and 32 units, rectified-linear activations and a final scalar output. It was trained to predict $\phi_k(a,b)$ using a Smooth L1 loss. Performance was evaluated on held-out data over 20 independent random 75\%/25\% train-test splits using Pearson correlation and mean squared error.

\subsubsection{Siamese energy model}
\label{sec:methods_siamese}

The Siamese model used two identical branches with shared weights. Each branch mapped one seven-base sequence to an energy $E_{\theta}(s)$, and the model output was the difference in Eq.~\eqref{eq:energy_model}. The base network was convolutional. The first convolutional layer had kernel size $r_{\mathrm{int}}$, which was varied from one to seven bases to control the interaction range. Subsequent $1\times1$ convolutions, rectified-linear activations and global average pooling produced the energy. Models were trained with a Smooth L1 loss on the empirical log-ratio. For each $r_{\mathrm{int}}$, performance was evaluated by 20-fold cross-validation.

\subsubsection{Chargaff-aware energy model}
\label{sec:methods_chargaff}

The Chargaff-aware model augmented each training example with reverse-complement sequences $(k_a^{\RC},k_b^{\RC})$. It predicted a symmetrized energy difference
\begin{equation}
\hat{y}_{\mathrm{main}}=\frac{1}{2}\left[E_{\theta}(k_a)+E_{\theta}(k_a^{\RC})\right]-\frac{1}{2}\left[E_{\theta}(k_b)+E_{\theta}(k_b^{\RC})\right]
\end{equation}
and an auxiliary reverse-complement difference
\begin{equation}
\hat{y}_{\mathrm{aux}}=E_{\theta}(k_a)-E_{\theta}(k_a^{\RC}).
\end{equation}
The loss was
\begin{equation}
L=\mathrm{SmoothL1}(\hat{y}_{\mathrm{main}},\phi_k(a,b))+\lambda\,\mathrm{SmoothL1}(\hat{y}_{\mathrm{aux}},0),
\end{equation}
with all branch weights shared. This model tests whether imposing reverse-complement symmetry improves generalization or stabilizes the inferred energy landscape.

\subsection{Cycle affinities and reproducibility}
\label{sec:methods_cycles}

For each fixed flank $k$, we considered closed three-node cycles among the four central bases and computed the affinity in Eq.~\eqref{eq:cycle}. A reversible field has $\mathcal{A}_k(a,b,d)=0$ for every closed cycle. There are four unordered three-base cycles per flank; the reported count of 16,384 uses one canonical orientation for each cycle, while the reverse orientation would give the same magnitude with the opposite sign. We summarized the distribution of signed and absolute affinities across all flanks and canonical cycles.

To test whether cycle affinities were shared across populations, we computed the one-way random-effects intraclass correlation coefficient. If $Y_{ip}$ denotes the cycle-affinity statistic for item $i$ (a flank together with a canonical three-base cycle) in population $p$ and there are $m=3$ populations, then
\begin{equation}
\mathrm{ICC}(1)=\frac{MS_{\mathrm{between}}-MS_{\mathrm{within}}}{MS_{\mathrm{between}}+(m-1)MS_{\mathrm{within}}},
\end{equation}
where $MS_{\mathrm{between}}$ is the mean square across cycle/flank items and $MS_{\mathrm{within}}$ is the residual mean square across populations within the same item.

\subsection{Biological annotation of irreversible cycles}
\label{sec:methods_cpg}

To assess whether high-affinity cycles were biologically interpretable, we annotated each three-base cycle by CpG content and substitution class. A cycle was classified as CpG-linked if at least one of its three seven-base states contained a CpG dinucleotide overlapping the central mutated base. We separately annotated CpG-transition edges, including C$\to$T in a CpG-compatible state and the reverse-complement-equivalent G$\to$A edge, as well as the opposite directions T$\to$C and A$\to$G. We also recorded whether the three-base cycle contained an A/G or C/T transition class and which base was excluded from the three-base cycle.

For each unique cycle and flank, we computed the mean $|\mathcal{A}|$ across African, Asian and European mutation probability estimates. High-curl cycles were defined as the top 10\% or top 5\% of this distribution. Enrichment of each annotation among high-curl cycles was quantified by Fisher's exact test, reporting odds ratios, high-curl fractions and background fractions. Differences in $|\mathcal{A}|$ between annotated and non-annotated cycles were tested using the Mann--Whitney U test. Because cycles sharing contexts or edges are not fully independent, these $p$ values are interpreted as descriptive support for the reported enrichments, whereas the main evidence is the magnitude and robustness of the odds ratios. Very small numerical $p$ values that underflowed to zero were reported as $p<10^{-300}$.

To quantify cross-population reproducibility of the signed curl signal, we performed leave-one-population-out prediction. For each population, the signed cycle affinity was predicted as the mean affinity of the other two populations, and compared with the held-out value using Pearson correlation, Spearman correlation, mean squared error and the slope through the origin.

\subsection{Statistics and implementation}
\label{sec:methods_statistics}

Reported neural-network performance values are means and standard deviations across cross-validation folds or independent train-test splits, as specified in the Results. Correlations are Pearson correlations. Model training, short-word aggregation, cycle-affinity analysis, Hodge projection and figure generation were implemented in the accompanying code repository.

\section{Data and code availability}

Context-dependent human mutation probability tables are available from Aggarwala and Voight \cite{Aggarwala2016}. Human reference genome sequences were obtained from GRCh38/hg38. Processed data used to reproduce the analyses, figures and numerical results, together with code for data processing, neural-network training, Hodge decomposition and figure generation, are available at \url{https://github.com/compbiomed-unito/Human_Mutation_Field}.

\begin{acknowledgments}
 I.C., D.M.B., A.M. and P.F. conceived the study. I.C., S.P. and P.F. performed data processing, neural-network modelling and genomic-composition analyses. D.M.B. and A.M. contributed the non-equilibrium statistical-physics framework and interpretation. All authors analyzed results, discussed the manuscript and approved the final version.
\end{acknowledgments}

\appendix

\section{Additional irreversible cycles and model performance}
\label{app:additional}

\begin{figure*}[t]
    \centering
    \figbox[width=\textwidth]{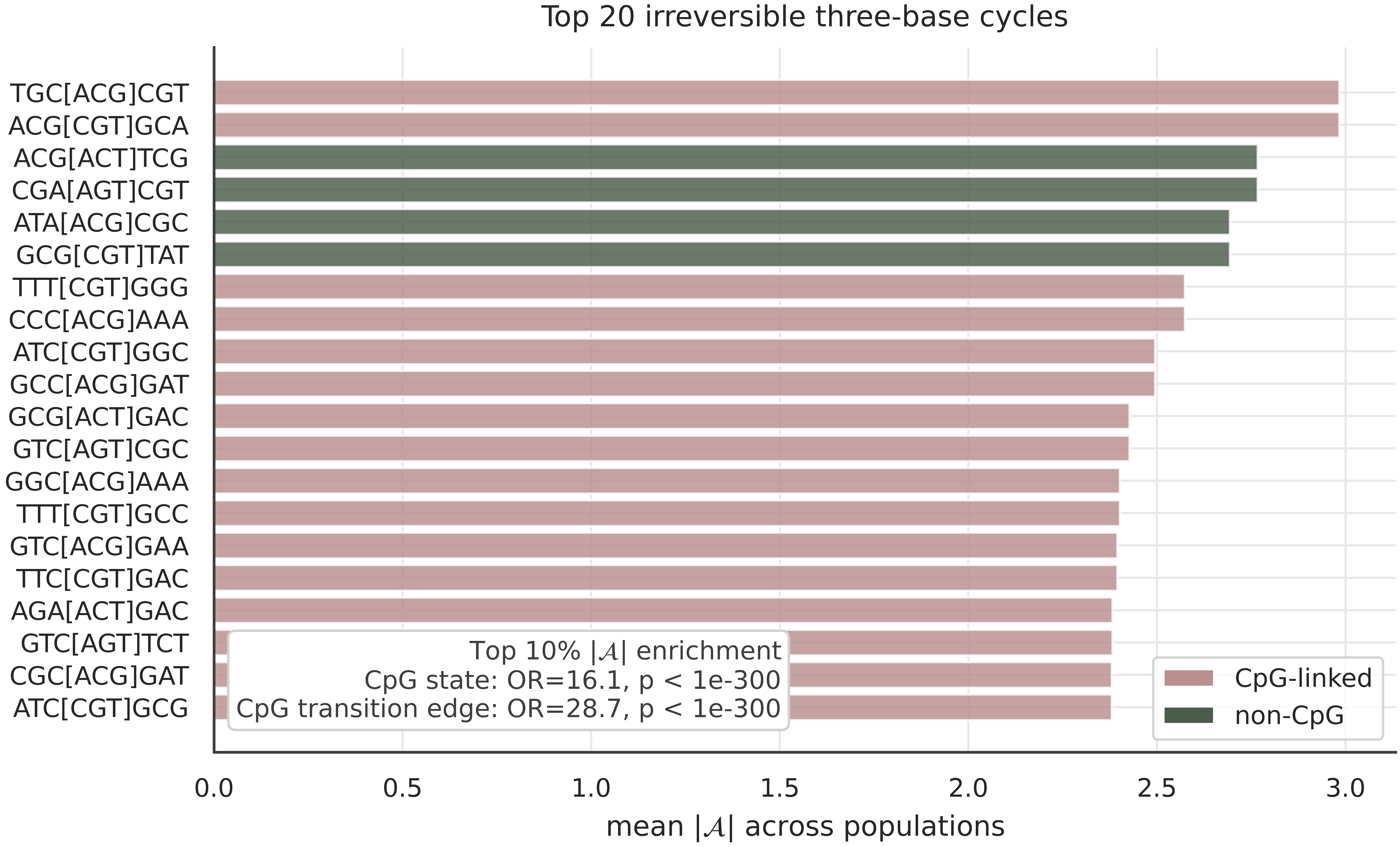}
    \caption{\textbf{Top irreversible three-base cycles.}
    The top 20 cycles ranked by mean absolute affinity, $\mathcal{A}$, across populations are shown. Labels indicate the left three-base flank, the central three-base cycle in brackets and the right three-base flank. CpG-linked cycles are shown in the pink palette and non-CpG cycles in the green palette. The enrichment summary highlights that the highest-affinity cycles are dominated by CpG-linked states and CpG-transition edges.}
    \label{fig:top_cycles}
\end{figure*}

\begin{table}[t]
\centering
\caption{\textbf{Predictive performance of the Siamese energy model across interaction lengths.} Pearson correlation and mean squared error are reported for each interaction range $r_{\mathrm{int}}$. These detailed values correspond to the performance curve shown in Fig.~\ref{fig:energy}.}
\label{tab:siamese_performance_by_k}
\begin{ruledtabular}
\begin{tabular}{ccc}
$r_{\mathrm{int}}$ & Pearson correlation & MSE \\
1 & $0.617 \pm 0.009$ & $0.872 \pm 0.019$ \\
2 & $0.890 \pm 0.005$ & $0.289 \pm 0.011$ \\
3 & $0.912 \pm 0.004$ & $0.234 \pm 0.007$ \\
4 & $0.924 \pm 0.003$ & $0.203 \pm 0.006$ \\
5 & $0.933 \pm 0.003$ & $0.182 \pm 0.007$ \\
6 & $0.933 \pm 0.003$ & $0.181 \pm 0.006$ \\
7 & $0.934 \pm 0.002$ & $0.178 \pm 0.005$ \\
\end{tabular}
\end{ruledtabular}
\end{table}

\section*{Conflicts of interest}

The authors declare no competing interests.

\bibliography{refs_arxiv}

\end{document}